\documentstyle[preprint, aps]{revtex}
\begin{document}

\draft

\date{October 17, 1997, accepted for publication January 1998}
\title{
Magnetoresistance due to Domain Walls in Micron Scale Fe Wires\\
with Stripe Domains
}
\author
{
A. D. Kent$^a$, U. Ruediger$^a$, J. Yu$^a$, S. Zhang$^a$, P. M. Levy$^a$ 
Y. Zhong$^b$, S. S. P. Parkin$^c$}

\address{
$^a$New York University, Department of Physics, 4 Washington Place,
New York, NY 10003
\\
$^b$Physics Department, City College of the City University of New York, 
NY 10031
\\
$^c$IBM Research Division, Almaden Research Center, San Jose, CA 95120
}
\maketitle

\begin{abstract}

The magnetoresistance (MR) associated with domain boundaries has
been investigated in microfabricated bcc Fe (0.65 to 20 $\mu$m
linewidth) wires with controlled stripe domains. 
Domain configurations have been characterized using MFM. 
MR measurements as a function of field angle, temperature and domain 
configuration are used to estimate MR contributions due to resistivity 
anisotropy and domain walls. 
Evidence is presented that domain boundaries enhance the conductivity in 
such microstructures over a broad range of temperatures (1.5 K to 80 K).

\end{abstract}

\newpage
\section{Introduction}
The low field magnetoresistance (MR) of a ferromagnetic metal depends in a
detailed and usually complex manner on its magnetic domain structure. 
An understanding of this interplay between transport and magnetic properties
is important to the miniaturization of magnetic devices to nanoscale
dimensions, such as those based on giant magnetoresistance (GMR), and to
the interpretation of transport results on magnetic nanostructures. 
For instance, a recent  experiment suggests large MR effects due to domain
walls can be observed even at room temperature in simple ferromagnet
films\cite{ref1}. The mechanism proposed is analogous to that operative in
GMR--the conductance in the presence of domain walls is reduced due a mixing
of minority and majority spin channels in the wall\cite{ref2,ref3}. 
Evidently, this might be exploited in novel magnetoelectronic devices and  
will effect the properties of conventional GMR devices as such devices are 
scaled down in size (so that walls may occupy an appreciably fraction of 
the device). Other research has focused on the use of magnetoresistance
as a probe of domain wall dynamics at low temperature in nanometer scale Ni 
wires\cite{ref4}. Here it appears that domain boundaries enhance the sample
conductance. A novel theoretical explanation has been suggested in which
domain walls destroy the electron coherence necessary for weak
localization\cite{ref6} at low temperature. This recent research points to the
need for experiments over a range of temperatures on microstructures with well characterized and controllable domain patterns to isolate the important
contributions to the MR in small samples.

Here we report on the first of such experiments. Fe microstructures with
stripe domains arranged perpendicular to the current direction
have been realized to study the effect of domain walls on magnetotransport
properties. In the following we discuss the fabrication, magnetic characteristics
and transport properties of these wires.

\section{Fabrication and Characterization}

The starting point for these experiments are expitaxial (110)
oriented bcc Fe thin films. These films have a large in-plane uniaxial
magnetocrystalline anisotropy, with the easy axis parallel to the [001]
direction. They are grown using an UHV e-beam evaporator on a-axis 
(11${\bar 2}$0) sapphire substrates. First a 10 nm thick (110) Mo seed
layer is deposited at a substrate temperature of 900 K followed by a 100 nm
thick layer of Fe at 510 K\cite{ref7}.  X-ray pole figures show that Fe (110)
layers grow with their in-plane [1${\bar 1}$1] axis parallel to the [0001] axis
of the sapphire substrate.A small ($< 0.05\%$) and anisotropic in-plane strain
is also found, consistent with previous x-ray studies of thinner films\cite{ref7}. 
The films are then patterned using projection optical lithography to produce
micron scale wires (0.65 to 20 $\mu$m linewidths) with the wire oriented
perpendicular to the magnetic easy axis. 
Fig. \ref{fig:1} shows the geometry of a 1 $\mu$m linewidth transport structure.

The competition between magnetocrystalline, exchange and magnetostatic
interactions in the wire results in a pattern of regularly spaced stripe domains. 
Varying the linewidth changes the ratio of magnetostatic to domain
wall energy and hence the domain size. The magnetic domain configuration is
also strongly affected by the magnetic history of the samples. 
Fig. \ref{fig:2} shows MFM images of the domain configuration of a 1.5 $\mu$m 
(Fig. \ref{fig:2}~a and b) and 20 $\mu$m wire (Fig. \ref{fig:2}~c and d)
in zero field with a vertically magnetized tip.These images highlight the domain
walls and magnetic poles at the wire edges.  Before performing these MFM
measurements the wires were magnetized to saturation with a magnetic field
transverse (Fig. \ref{fig:2}~a and c) or longitudinal (Fig. \ref{fig:2}~b and d)
to the wire axis. For the 1.5 $\mu$m wire in the
transverse case the average domain width is 1.7 $\mu$m and much 
larger than in the longitudinal case, where the average domain width is
0.4 $\mu$m.

In Fig. \ref{fig:3} the domain width is plotted as a function of wire width.  
In all samples closure domains of a triangular shape are found near the wire
edges. White dotted lines in Fig. \ref{fig:2}~a) illustrate the approximate domain
structure. In order to determine the MR contributions due to resistivity
anisotropy the volume fraction of closure domains (with {\bf M} $\parallel$ 
{\bf J}) must be estimated. Fig. \ref{fig:3} also shows this fraction (labeled $\gamma$) determined from  MFM images after magnetic saturation in either the transverse or longitudinal direction.

\section{Transport Properties}

\vspace{-0.25cm}
\subsection{Results}
MR measurements were performed in a variable temperature high field
cryostat with in-situ (low temperature) sample rotation capabilities.  
The applied field was always in the plane of the thin film, oriented either
longitudinal or transverse to the wire axis. The magnetic history of the
sample was carefully controlled and low ac current levels were used 
($\sim$10 $\mu$A). Fig. \ref{fig:4} shows representative results at 40 K
on two samples with distinct domain configurations; 
Fig. \ref{fig:4}~a), b) a 1.5 $\mu$m wire  and Fig. \ref{fig:4}~c), d) 
a 20 $\mu$m wire.  There is structure to the MR in applied fields less than
the saturation field, after which the resistivity increases monotonically
with field. Fig. \ref{fig:4} also shows that the 1.5~$\mu$m wire resistivity
depends on its magnetic history, being larger after saturation in the transverse
geometry.

The MFM images of domain configurations are performed at room temperature.
Thus to correlate low temperature MR measurements with domain
configurations we warm the sample to room temperature, cycle the magnetic
field to establish a known ${\bf H}=0$ magnetic state, and cool. 
The resistivity at ${\bf H}=0$ and the MR are unchanged for this sample in both longitudinal and transverse measurement geometry\cite{ref8}. 
This is strong evidence that the domain structure is not affected by 
temperature in this range and consistent with SQUID hysteresis loop 
measurements\cite{ref9}.

\subsection{Interpretation}
There are two important sources of low field low temperature MR which must
be considered to interpret this transport data. The first has its origins
in spin-orbit coupling\cite{ref10} and is known as anisotropic magnetoresistance
(AMR)--the resistivity extrapolated back to zero internal field ($B=0$)
depends on ${\bf M} \cdot {\bf J}$. The second effect is due to the ordinary (Lorentz)
magnetoresistance and also in general anisotropic ({\it i.e.} dependent on 
${\bf J} \cdot {\bf B}$). As Fe has a large magnetization and hence a large internal magnetic field (4$\pi M =2.2$ T) both factors are of importance.  
The resistivity of domains parallel and perpendicular to the current direction can be written as:
\begin{eqnarray}
\rho_{\perp}(B,T) = \rho_{\perp}(0,T)[1+F_{\perp}(B/\rho_{\perp}(0,T))]
\\
\rho_{\parallel}(B,T) = \rho_{\parallel}(0,T)[1+F_{\parallel}(B/\rho_{\parallel}(0,T))]
\end{eqnarray}
here $B$ is the internal field in the ferromagnet; 
$B=4\pi M+H-H_d$, 
with $H$ the applied field and $H_d$ the demagnetization field. The AMR is proportional to $\rho_{\parallel}(0,T)-\rho_{\perp}(0,T)$. The function $F$ is know as the Kohler function and parametrizes the ordinary magnetoresistance for longitudinal and transverse
field geometry in terms of  $B/\rho \sim \omega_c \tau$.  With this form we have been able to determine the scaling functions $F_{\perp}$ and $F_{\parallel}$\cite{ref11} and 
$\rho_{\perp}(0,T)$ and $\rho_{\parallel}(0,T)$. 
At 40 K the AMR is $\sim$$10^{-3.}$. 
The fact that $\rho_{\perp}(H) > \rho_{\parallel}(H)$ in Fig. \ref{fig:4} 
(and thus that the resistivity anisotropy is opposite in sign to the AMR in this range of B
fields) is a consequence of the ordinary MR ({\it i.e.} $F_{\perp}^\prime >
F_{\parallel}^\prime $)\cite{ref9}.

The low field MR has been analyzed as follows. As this resistivity
anisotropy is small and the domain size much greater than the mean free
path we can write the effective resistivity in the $H=0$ magnetic state 
as\cite{ref12}:
\begin{equation}
\rho_{eff}(H=0,T) = 
\gamma \rho_{\parallel}(B_i,T) + (1-\gamma)\rho_{\perp}(B_i,T)
\end{equation}
where $\gamma$ is the volume fraction of domains oriented longitudinally and 
$B_i$ isthe field internal to these domains  ($=4\pi M-H_d$). 
We determine $\rho_{\perp}(B_i,T)$  and $\rho_{\parallel}(B_i,T)$  by
extrapolation of the MR data above saturation (as indicated by the dashed
lines in Fig. \ref{fig:4}). 
With this model we estimate the low field MR associated with domain rotation 
and the sources of resistivity anisotropy discussed above. 
In the longitudinal geometry, at $H=0$ the 1.5~$\mu$m wire consists
predominately of domains perpendicular to ${\bf J}$ ($\gamma = 0.18$). 
Thus  $\rho_{eff}$ is seen to be larger than that measured (see Fig. \ref{fig:4}b). 
In the transverse case the measured resistivity is close to the estimation of $\rho_{eff}$. 
In the transverse case the domain wall density is also 4 times smaller. 
For the 20 $\mu$m sample in both orientations the MR is consistent with 
estimations.

The deviations from this model $\rho_d=\rho(H=0)-\rho_{eff}(H=0)$,
 i.e., the measured H=0 resistivity
minus the effective resistivity, are negative and depend systematically on
domain wall density, increasing in magnitude with increasing domain wall
density\cite{ref9}. We also find that $|\rho_d|$  decreases with increasing
temperature approaching zero at $\sim$80 K (Fig. \ref{fig:5}) and above. 
These deviations appear consistent with an enhancement of the conductivity 
associated with domain walls in these wires at low temperatures.
In contrast to reports on Co and Ni films \cite{ref1}\cite{ref2}, increasing
resistivity due to domain walls at room temperature 
is not observed within the experimental uncertainty of
$\rho_d/\rho(H=0) \sim 10^{-3}$.

In summary, we have established a means of realizing magnetic microstructures 
with controlled domain patterns enabling new experimental studies of the role 
of domain boundaries on transport. The Lorentz MR is found to be important in
determining the resistivity  anisotropy and magnetotransport properties in these Fe wires. 
Our initial results also suggest that the presence of domain boundaries enhances
the sample conductance. While an enhancement is consistent with a recent theory based on weak
localization, the magnitude of the effect appears larger than predicted\cite{ref6}
and the effect is present up to~$\sim$80~K. These magnetic microstructures will certainly
 permit more detailed investigations of these novel effects.

\section*{Acknowledgement}
This research is supported by DARPA-ONR Grant\# N00014-96-1-1207.
Y. Z. is supported by AFOSR Grant\# F49620-92-J-0190. We thank C. Noyan for x-ray characterization, H. Zhang and B. Sinkovic for help with MOKE,
M. P. Sarachik for SQUID measurements and M.Ofitserov for technical
assistance. Microstructures were prepared at the CNF, project \#588-96.


\begin{figure}
\caption{
Optical micrograph of 1 $\mu$m transport structure showing the Fe
crystallographic orientation.
}
\label{fig:1}
\end{figure}

\begin{figure}
\caption{
MFM domain images in zero applied field for a) and b) a 1.5 $\mu$m and
c) and d) a 20 $\mu$m linewidth wire.  In a) and c) the wire is first
magnetized transverse to the wire axis.  In b) and d) the wire is first
saturated along the wire axis.
}
\label{fig:2}
\end{figure}

\begin{figure}
\caption{
Heavy solid curve right hand axis: domain size along the wire axis
after either transverse (solid squares) or longitudinal (solid circles)
saturation versus linewidth. Volume fraction of longitudinally oriented
closure domains after transerve (open squares) and longitudinal (open
circles) saturation versus linewidth.
}
\label{fig:3}
\end{figure}

\begin{figure}
\caption{
MR data at 40 K of a 1.5 $\mu$m wire in a) a transverse and b) a
longitudinal geometry $\rho_{\perp}$(H=0,40 K)=1.2 $\mu$$\Omega$cm.
Results from a 20 $\mu$m wire are
shown in parts c) and d) $\rho_{\perp}$(H=0,40 K)=0.24 $\mu$$\Omega$cm.
}
\label{fig:4}
\end{figure}

\begin{figure}
\caption{
$\rho_d$  as a function of temperature for a 1.5 $\mu$m transport line.
}
\label{fig:5}
\end{figure}


\begin{references}

\bibitem{ref1}
J. F. Gregg, et. al.,  ``Giant magnetoresistance effects in a single
element magnetic thin film,'' {\it Phys. Rev. Lett.}, vol. 77, pp. 1580-1583, August
1996.

\bibitem{ref2}
M. Viret, et. al.,  ``Spin scattering in ferromagnetic thin films,'' 
{\it Phys. Rev B},  vol. 53, pp. 8464-8468, April 1996.

\bibitem{ref3}
P. M. Levy and S. Zhang, ``Resistivity due to domain wall scattering,''
{\it Phys. Rev. Lett.}, vol. 79, pp. 5110-5113, Dec. 1997.

\bibitem{ref4}
K. Hong and N. Giordano,  ``Approach to mesoscopic magnetic measurements,''
{\it Phys. Rev. B}, vol. 51, pp. 9855-9862, April 1995 and K. Hong and N. Giordano,
preprint 1997.

\bibitem{ref6}
G. Tatara and H. Fukuyama, ``Resistivity due to a domain wall in 
ferromagnetic metal,'' {\it Phys. Rev. Lett.}, vol. 78, pp. 3773-3776, May 1997.

\bibitem{ref7}
B. M. Clemens, et. al., ``In-situ observation of anisotropic strain 
relaxation in expitaxial Fe (110) films on Mo (110),'' 
{\it J. of Mag. and Mag. Mat.}, vol. 121, pp. 37-41 Jan 1993.

\bibitem{ref8}
In the case of  a 0.65 $\mu$m wire there were noticable differences in MR
after this temperature and field cycling.

\bibitem{ref9}
U. Ruediger, J. Yu, S. Zhang, A. D. Kent, and S. S. P. Parkin, 
``Negative domain wall contribution to the resistivity of 
microfabricated Fe wires,'' to be published (1998).

\bibitem{ref10}
see, for example, I. A. Campbell and A. Fert, "Transport Properties of
Ferromagnets," in {\it Ferromagnetic Materials vol. 3}, E. P. Wohlfarth, Ed.,
North-Holland Pub. Co. 1982, pp. 747-806.

\bibitem{ref11}
F. C. Schwerer and J. Silcox,  ``Electrical resistivity of Nickel at low
temperatures,'' {\it Phys. Rev. Lett.}, vol. 20, pp. 101-103, January 1968.

\bibitem{ref12}
Since the current density in each domain is to a good approximation
independent of the domain configuration this weighted average is
appropriate.

\end{references}
\end{document}